\documentclass[conference]{IEEEtran}
\IEEEoverridecommandlockouts
\usepackage{cite}
\usepackage{url}
\usepackage{amsmath,amssymb,amsfonts}
\usepackage{algorithmic}
\usepackage{graphicx}
\usepackage{textcomp}
\usepackage{xcolor}
\def\BibTeX{{\rm B\kern-.05em{\sc i\kern-.025em b}\kern-.08em
    T\kern-.1667em\lower.7ex\hbox{E}\kern-.125emX}}
\begin{document}

\title{Multi-Agent eXperimenter (MAX)}

\author{
    \IEEEauthorblockN{Önder Gürcan}
    \IEEEauthorblockA{
        \textit{Université Paris-Saclay, CEA, List} \\
        Palaiseau, France \\
        onder.gurcan@cea.fr
    }
}

\maketitle

\begin{abstract}
We present a novel multi-agent simulator named Multi-Agent eXperimenter (MAX) that is designed to simulate blockchain experiments involving large numbers of agents of different types acting in one or several environments.
The architecture of MAX is highly modular, enabling easy addition of new models.
\end{abstract}


\section{Introduction}

We live in an increasingly complex and connected world where individuals and organizations need to better understand how they are impacted by a wide range of potential scenarios.
Traditional methods of modeling do not fit for purpose since they do not capture the complexity of real-world systems and thus they fail to effectively harness the huge power of technology and big data. 

In this paper, we identify that there is a gap to create a technology solution that could model blockchain systems and create a virtual environment that could enable  dynamic testing\footnote{Dynamic testing is carried out with executing the code and validating the output with the expected outcome. Dynamic testing is performed at all levels of testing and it can be either black or white box testing.} of hypothesis. 
Then we conclude that computer simulations\footnote{Computer simulation is the discipline of designing a model of an actual or theoretical system, executing the model on a digital computer, and analyzing the execution output \cite{Fishwick1997}.} is the key to radically better decisions.
Based on this conclusion, we propose a novel agent-based simulator called Multi-Agent eXperimenter (MAX)\footnote{https://cea-licia.gitlab.io/max/max.gitlab.io/} since agent-based modeling and simulation (ABMS) is universally recognized as the solution to modeling complex, dynamic and open systems. 
ABMS models and simulates the actions and interactions of individuals, organizations, groups and/or other entities (by abstracting them as agents) so we can assess their impact.

MAX\footnote{See MAX website: \url{https://cea-licia.gitlab.io/max/max.gitlab.io/}} is CEA tool that implements an advanced blockchain simulation framework that brings multi-agents
to distributed systems.
MAX (for ‘Multi-Agent eXperimenter’) is a framework for agent-based simulation and full agent-based
applications building. It was started by the author of this paper in 2017 (first commit in August 2017) \cite{Gurcan2019,Gurcan2020} and has
been in continuous development ever since at the CEA \cite{Lagaillardie2019,Mahe2024}. 
The objective of MAX is to make rapid prototyping
of industrial cases and to carry out their feasibility analysis in a realistic manner.
MAX is en open-source project under the EPL-2.0 license. MAX was used in the REACH\footnote{\url{https://www.reach-incubator.eu}, last access on 12/04/2024.} and ADACORSA\footnote{\url{https://adacorsa.eu}, last access on 12/04/2024.} European projects, and the Blockchain for Nuclear \cite{Mahe2024} industrial project in collaboration with EDF R\&D and Framatome.

MAX provides the computational abstractions necessary to model and simulate agent-based simulations for  systems at all levels since it follows an \textit{organization centered} modeling approach. 
This way, it is possible to model and simulate complex experiments involving large numbers of agents of different types acting in one or several systems.

Moreover, MAX offers the ability to better test new algorithms without much effort thanks to its modular and generic modeling approach.
This way, new complex systems (like blockchain systems) can be modeled with less effort and also applications that aim to be developed on top of a blockchain system can be easily tested on different blockchain systems.

The contributions of the article are as follows:

\begin{itemize}
\item We propose a novel agent-based simulator called Multi-Agent eXperimenter (MAX) 
   \begin{enumerate} 
   \item which better captures and models the requirements of blockchain systems,
   \item whose architecture is based the Agent/Group/Role (A/G/R) organization model,
   \item that has an integrated dynamic testing capability allowing it to build a modular and reusable model library that can be continuously integrated and delivered,
   \end{enumerate}
\item MAX provides a dedicated and modular Generic Blockchian model\footnote{https://gitlab.com/cea-licia/max/models/ledgers/max.model.ledger.blockchain} that 
   \begin{enumerate}
   \item  which is based on a blockchain abstract datatype formalization,
   \item  serves as a high-level software blueprint for any type of blockchain system and thus can easily be used for building new blockchain models,
   \item  allows us to evaluate design and implementation decisions about different blockchain systems both independently and as an ecosystem,
   \end{enumerate}
\end{itemize}

\section{Overall Architecture of MAX}
\label{sec:MAX}

The basic elements of ABMS are \textit{agents}, \textit{simulated environments} and the \textit{simulation environment} \cite{Klugl2005}.
The \textit{simulation environment} (or infrastructure) is the underlying environment for executing simulation models. 
Independent from a particular model, it controls the specific simulation time advance and
scheduling of actions, and carry the global state variables that affect all the agents situated in it.

\textit{Agents} are proactive and autonomous entities that try to fulfill their goals by interacting with other agents and/or simulated environments in which they are situated. 
To better capture and model the requirements of blockchain systems, MAX follows an \textit{organization centered} approach rather than an \textit{agent centered} one. 
Organizations represent complex entities where a multitude of agents interact, within a structured environment aiming at some global purpose \cite{Dignum2009}.
For modeling, designing and implementing  organizations, the key concepts are roles, groups, tasks, interaction protocols and norms (or rules).  
In a simple yet elegant organization centered multi-agent approach proposed by  \cite{Ferber2003} (the Agent/Group/Role model), an organization is partitioned into one or several \textit{groups} that can overlap and is composed of \textit{agents} playing one or several \textit{roles} that are functionally related to the group(s) they belong to.   
Each \textit{group} constitutes a context of interaction of agents.
Agents belonging to a group may discover each other and can communicate using the same language and protocol.
A \textit{role}, on the other hand, is an abstract representation of a functional position of an agent in a group.
Roles are local to groups and can be played by several agents.

A \textit{simulated environment} contains agents and non-agent entities of the simulation model. 
It can also carry some state variables that affect all 
the agents situated in it and can have its own dynamics like the creation of a new agent or coordinating the interactions between agents.
Both the computer science community \cite{Weyns2007} and the fundemantal sciences community \cite{Pinter-Wollman2017} agree that \textit{environments} have considerable impact on the behavior of agents and thus should be \textit{first class abstractions} when modeling multi-agent systems.
Environments can be physical (e.g., physical network infrastructure) as well as logical (e.g., social network) and are usually modeled through the resource abstraction as a non-goal driven entity producing events and/or waiting for requests to perform its function. 
Environments are essential in \textit{coordination} since they work as mediators for agent interaction through which agents can communicate and coordinate \textit{indirectly}.
In other words, agents never interact directly and all interactions are mediated through environments.
Environments feature autonomous dynamics and affect agent coordination.
Furthermore, environment is also important if we want to use computational approaches to learning from interaction such as reinforcement learning \cite{Sutton2018} since it allows to model how the environment responds to what the agents do. 

Several extensions of the A/G/R organization have been proposed to integrate \textit{environment} before \cite{Odell2002, Ferber2004}.
However, in all of them, \textit{environment} is an extension to the A/G/R model.
Conversely, in our study, we chose to unify the \textit{group} and the \textit{environment} abstractions since \textit{environment} is also a context of interaction of agent, and we define it as an \textit{agent} since it features autonomous dynamics and mediates agent interactions\footnote{In other words, environment is an agent that plays the role "Environment" (or "Mediator" or "Coordinator").}. 
Consequently, the organization model we propose is called the Agent/Group/Role (A/G/R) model.

To define a new simulation model in MAX, the A/G/R organization model should be followed by using elements given in the modelling level. 

\section{Discussion and Conclusions}

MAX is a modular agent-based simulation framework based on the Agent/Environment/Role model  designed to simulate complex experiments involving large numbers of agents playing different roles in one or several blockchain systems.
We observed that the performance overhead brought by such modularity is acceptable.
However, we have not conducted detailed performance analysis to come to a concrete conclusion.
On the contrary, thanks to the modularity, the execution of simulations can increase precision and reduce the time spent for the analyses.
One can quickly model and develope her/his new \textit{blockchain} model by reusing the existing model elements that have already been modeled, developed, tuned and tested. 

As highlighted in the literature, one major challenge for the modeling and simulation community is to find ways of sharing modeling specifications in an efficient way, enabling the replication, and thus the study of published experiments \cite{Axelrod1997}.
Otherwise, it may lead to a problem which is identified as the \textit{engineering divergence phenomenon} in \cite{Michel2004,Edmonds2003} where different implementations of a particular conceptual model may give different outputs.
In this sense, replicating the simulation model on different simulation environments is proposed as a solution in some studies \cite{Wilensky2007,Sansores2005}.
MAX avoids this phenomenon thanks to its modular and reusable modeling architecture.

From the blockchain systems side, MAX provides a Generic Blockchain model which serves as a high-level software blueprint for any type of blockchain system. 
This allows any simulation scientist that wants to study any aforementioned phenomenon (i.e. distributed system, social science, economy and software engineering) to build easily (and without much effort) his/her dedicated blockchain model.
Any improvement in the Generic Blockchain model and/or in dedicated blockchain models are continuously calibrated and integrated as a whole. 

\bibliographystyle{ieeetr}
\bibliography{references}

\begin{thebibliography}{10}

\bibitem{Fishwick1997}
P.~A. Fishwick, ``Computer simulation: Growth through extension,'' {\em Trans.
  Soc. Comput. Simul. Int.}, vol.~14, p.~13–23, Mar. 1997.

\bibitem{Gurcan2019}
{\"{O}}.~G{\"{u}}rcan, ``Multi-agent modelling of fairness for users and miners
  in blockchains,'' in {\em 2nd Workshop on Block Chain Technologies 4
  Multi-Agent Systems (BCT4MAS), co-located with PAAMS 2019, Avila, Spain, June
  26-28, 2019}, 2019.

\bibitem{Gurcan2020}
O.~G\"{u}rcan, ``On using agent-based modeling and simulation for studying
  blockchain systems,'' in {\em JFMS 2020 - Les Journées Francophones de la
  Modélisation et de la Simulatio - Convergences entre la Théorie de la
  Modélisation et la Simulation et les Systèmes Multi-Agents}, (Cargèse,
  France), 2020.

\bibitem{Lagaillardie2019}
N.~Lagaillardie, M.~A. Djari, and O.~G\"{u}rcan, ``A computational study on
  fairness ofthe tendermint blockchain protocol,'' {\em Information}, vol.~10,
  no.~12, 2019.

\bibitem{Mahe2024}
E.~Mahe, R.~Abdallah, S.~Tucci-Piergiovanni, and P.-Y. Piriou,
  ``Adversary-augmented simulation to evaluate fairness on hyperledger
  fabric,'' 2024.

\bibitem{Klugl2005}
F.~Kl{\"u}gl, M.~Fehler, and R.~Herrler, ``About the role of the environment in
  multi-agent simulations,'' in {\em Environments for Multi-Agent Systems}
  (D.~Weyns, H.~Van Dyke~Parunak, and F.~Michel, eds.), vol.~3374 of {\em
  LNCS}, pp.~127--149, Springer Berlin / Heidelberg, 2005.

\bibitem{Dignum2009}
V.~Dignum, ``{The Role of Organization in Agent Systems},'' in {\em Handbook of
  Research on Multi-Agent Systems}, pp.~1--16, IGI Global, 2009.

\bibitem{Ferber2003}
J.~Ferber, O.~Gutknecht, and F.~Michel, ``{Agent/Group/Roles: Simulating with
  Organizations},'' in {\em {ABS'03: Agent Based Simulation}} (M.~J.P., ed.),
  (Montpellier (France)), p.~12, Apr. 2003.
\newblock April 28-30.

\bibitem{Weyns2007}
D.~Weyns, A.~Omicini, and J.~Odell, ``Environment as a first class abstraction
  in multiagent systems,'' {\em Autonomous Agents and Multi-Agent Systems},
  vol.~14, pp.~5--30, Feb 2007.

\bibitem{Pinter-Wollman2017}
N.~Pinter-Wollman, A.~Penn, G.~Theraulaz, and S.~M. Fiore, ``Interdisciplinary
  approaches for uncovering the impacts of architecture on collective
  behaviour,'' {\em Philosophical Transactions of the Royal Society of London
  B: Biological Sciences}, vol.~373, no.~1753, 2018.

\bibitem{Sutton2018}
R.~S. Sutton and A.~G. Barto, {\em Reinforcement Learning: An Introduction}.
\newblock Cambridge, MA, USA: A Bradford Book, 2018.

\bibitem{Odell2002}
J.~J. Odell, H.~Van Dyke~Parunak, M.~Fleischer, and S.~Brueckner, ``Modeling
  agents and their environment,'' in {\em Proceedings of the 3rd International
  Conference on Agent-Oriented Software Engineering III}, AOSE’02, (Berlin,
  Heidelberg), p.~16–31, Springer-Verlag, 2002.

\bibitem{Ferber2004}
J.~Ferber, F.~Michel, and J.~Baez, ``Agre: Integrating environments with
  organizations,'' in {\em Proceedings of the First International Conference on
  Environments for Multi-Agent Systems}, E4MAS’04, (Berlin, Heidelberg),
  p.~48–56, Springer-Verlag, 2004.

\bibitem{Axelrod1997}
R.~Axelrod, ``Advancing the art of simulation in the social sciences,'' in {\em
  Simulating Social Phenomena} (R.~Conte, R.~Hegselmann, and P.~Terna, eds.),
  (Berlin, Heidelberg), pp.~21--40, Springer Berlin Heidelberg, 1997.

\bibitem{Michel2004}
F.~Michel, A.~Goua{\"i}ch, and J.~Ferber, ``Weak interaction and strong
  interaction in agent based simulations,'' in {\em Multi-Agent-Based
  Simulation III} (D.~Hales, B.~Edmonds, E.~Norling, and J.~Rouchier, eds.),
  (Berlin, Heidelberg), pp.~43--56, Springer Berlin Heidelberg, 2003.

\bibitem{Edmonds2003}
B.~Edmonds and D.~Hales, ``{Replication, Replication and Replication: Some Hard
  Lessons from Model Alignmen},'' {\em Journal of Artificial Societies and
  Social Simulation}, vol.~6, no.~4, pp.~1--11, 2003.

\bibitem{Wilensky2007}
U.~Wilensky and W.~Rand, ``Making models match: Replicating an agent-based
  model,'' {\em Journal of Artificial Societies and Social Simulation},
  vol.~10, no.~4, p.~2, 2007.

\bibitem{Sansores2005}
C.~Sansores and J.~Pavon, ``Agent-based simulation replication: A model-driven
  architecture approach,'' in {\em 4th Mexican Internation Conference on
  Artificial Intelligence (MICAI'2005)}, (Mexica), pp.~244--253, 2005.

\end{thebibliography}

\end{document}